\documentclass[12pt]{article}
\usepackage{graphicx}
\newcommand{\odra}{O.~Dragoun}

\newcommand{\rys}{M.~Ry\v sav\'y}
\newcommand{\spa}{A.~\v Spalek}

\newcommand{\kas}{J.~Ka\v spar}
\begin{document}
\def\der#1#2{\frac{{\rm d}#1}{{\rm d}#2}}
\def\up#1{$^{#1}$}
\def\dn#1{$_{#1}$}
\def\mi#1{$-#1$}
\def\mgp#1{\marginpar{$\Leftarrow$#1}}
\def\mean#1{$<$$#1$$>$}
\def\Ttot{T_{\rm tot}}
\def\sigi{\sigma_{\rm init}}
\def\Szer#1{S(T_i;A,\Gamma_0,#1,N_b)}
\def\Srand#1{S_{rand}(T_i;A,\Gamma_0,#1,N_b)}

\newcommand{\beq}{\begin{equation}}
\newcommand{\eeq}{\end{equation}}
\newcommand{\bec}{\begin{center}}
\newcommand{\eec}{\end{center}}
\newcommand{\ben}{\begin{enumerate}}
\newcommand{\een}{\end{enumerate}}
\newcommand{\beqar}{\begin{eqnarray}}
\newcommand{\eeqar}{\end{eqnarray}}
\newcommand{\bit}{\begin{itemize}}
\newcommand{\eit}{\end{itemize}}
\newcommand{\plm}{$\pm$}
\newcommand{\plut}{$^{241}$Pu}
\newcommand{\ikry}{$^{83m}$Kr}
\newcommand{\kry}{$^{83}$Kr}
\newcommand{\alf}{$\alpha$}
\newcommand{\bet}{$\beta$}
\newcommand{\cob}{$^{57}$Co}
\newcommand{\gam}{$\gamma$}
\newcommand{\micron}{$\mu$m}
\newcommand{\Lea}{$\Leftarrow$}
\newcommand{\Ria}{$\Rightarrow$}
\newcommand{\sigmnusq}{$\sigma_{m_\nu^2}$}
\newcommand{\colden}{$\rho d$}
\newcommand{\asi}{$\sim$}
\newcommand{\mnu}{$m_\nu$}
\newcommand{\mnusq}{$m_\nu^2$}
\newcommand{\dmsq}{$\delta m_\nu^2$}
\newcommand{\krat}{$\times$}
\newcommand{\chisq}{$\chi^2$}
\newcommand{\dE}{\,{\rm d}E}
\newcommand{\KAT}{KATRIN}
\newcommand{\hzer}{\textsl{H\dn{0}}}
\newcommand{\hone}{\textsl{H\dn{1}}}
\newcommand{\ezer}{E_0}
\newcommand{\epri}{E_0^{'}}
%
\begin{flushright} Report NPI ASCR \v Re\v z\\ TECH--04/2005\vspace{3ex}\\
\end{flushright}
\bec \Large \bf
    Test of potential homogeneity in the KATRIN
    gaseous tritium source
\eec
%
%
\bec
    \rys\footnote{e-mail: rysavy@ujf.cas.cz}\\
   \small Nuclear Physics Institute, Acad. Sci. Czech Rep., \\
   \small CZ--250 68 \v{R}e\v{z} near Prague, Czech Republic
\eec
%
%
%
\date{}
\begin{abstract}
\ikry\ is supposed to be used to study the properties of the
windowless gaseous tritium source of the experiment KATRIN. In
this work we deduce the amount of \ikry\ which is necessary to
determine possible potential inhomogeneities via
conversion-electron-line broadening.
\end{abstract}
%
%
\section{Introduction}   \label{s:intro}
In the windowless gaseous tritium source (WGTS) for the experiment
\KAT, there should be -- in the ideal case -- a homogeneous
electrostatic potential which does not depend on time during the
exposure in one spectrum point. In fact, this does not need to be
so. It might be both inhomogenous (due to, e.g., charging caused
by the tritium decay or change of source's wall work function
along the WGTS) and time dependent. In this work, however, we
consider the inhomogeneity only.
\section{General}
To examine the inhomogeneity of the potential, we insert some
source of monoenergetic electrons into the WGTS. In particular, we
use gaseous \ikry. There exists a highly
converted\footnote{$\alpha_{tot}$=2011 -- evaluated by program
\cite{Ry77} using the atomic potential of \cite{Lu71}} isomeric E3
transition with the energy $E_\gamma$=32151.5~eV \cite{Pic92}. The
conversion electrons from the L\dn{3}-subshell which are most
suitable for our purpose have the kinetic energy $E_0$=30473.1~eV
(using the binding energy of \cite{Sev79}), the line width
$w$=1.19~eV \cite{Cam01}, and the corresponding relative
conversion coefficient is $\alpha_{L_3}/\alpha_{tot}$=0.381
\cite{Ry77}.

If the conversion electrons are being born in a region of
inhomogeneous potential, their registered kinetic energies vary.
This leads to line broadening and  also to a line shift. Our task
is to find the minimum quantity of \ikry\ needed to detect
potential inhomogeneities at the level of \asi 50~mV.

Let us imagine that the WGTS is cut into slices -- as a salami --
of the widths of d$x$. Then, at the position $x$, there is a
potential distorsion $\delta_U(x)$ and the tritium gas density
$\rho(x)$. Let the probability that the electron undergoes $i$
non-elastic scatterings prior to exit the WGTS be $p_i(x)$. Taking
into account the small width of the studied L\dn{3} line and the
finiteness of the electron energy loss in one scattering we can
neglect all $p_i$'s except $p_0$.

The conversion line is described by a Lorentzian function. Then
the count rate of the conversion electrons born in the interval
$\langle x,x+$d$x)$ is
\beq \label{e:elemK}
          K(E;E_0,w,x)\ \textrm{d}x = \frac{C}{2\pi}\  \frac{w \rho(x)
         p_0(x)}{\left[ E-E_0-e\delta_U(x)\right]^2 +(w/2)^2} \ \textrm{d}x.
\eeq
It is easy to find that the correct normalizing constant, $C$, is
given by
\beq
      C = \frac{A\times \alpha_{L_3}/\alpha_{tot}}{\int_0^L{\rho(x)\
      \textrm{d}x}},
\eeq
where $A$ is the \ikry\ activity in the total volume of WGTS and
integration runs over the whole length of WGTS.

To achieve the total L\dn{3} count rate, the formula
(\ref{e:elemK}) must be integrated for x running over the whole
WGTS length and we get
\beq \label{e:diff}
        K(E;E_0,w) = A\times\frac{\alpha_{L_3}}{\alpha_{tot}}
        \times\frac{w}{2\pi}\times\int_0^L{\frac{w \rho(x)
         p_0(x)}{\left[ E-E_0-e\delta_U(x)\right]^2 +(w/2)^2} \
         \textrm{d}x} / \int_0^L{\rho(x)\
      \textrm{d}x}.
\eeq
The KATRIN spectrometer is an integrating one. To express the
shape which will be registered, the differential form
(\ref{e:diff}) must be convoluted with the instrumental response
function \cite{LoI}. This depends on the magnetic field
intensities in the analyzing plane ($B_A$), at the source position
($B_s$), and the maximum intensity ($B_{max}$). Finally, we must
multiply the result by the reduction factor due to the
spectrometer acceptance angle \cite{LoI}
\beq
      R_{acc} = \frac{1}{2}(1-\cos{\vartheta_{max}})
\eeq
where $\sin{\vartheta_{max}} = \sqrt{B_s/B_{max}}$.
\section{Method}
In the experiment, we can detect the existence of inhomogeneities
by measuring the krypton line with and without, respectively, the
presence of tritium in the WGTS and then mutually comparing the
spectra. (No tritium implies no charging and then the
inhomogeneities are at least suppressed). Moreover, utilizing the
fact that the detector is position sensitive (pixel-like), we can
analogically search for different inhomogeneities in various parts
of WGTS by mutually comparing spectra taken by different pixel
groups. (Characterizing the inhomogeneities by some quantity
$\Delta U$ -- see below -- we thus can distinguish if there are
different $\Delta U$'s in different WGTS parts.) For the spectra
comparison, we apply the statistical tests \cite{Dra97}.

To find in advance the sensitivity of this procedure, we apply the
tests on simulated spectra corresponding to various kinds of
inhomogeneities. Repeating such a procedure many times supplies a
reasonable estimate of sensitivity.

\begin{figure}[h]
  \centerline{\includegraphics[clip=on,angle=0,width=10cm]{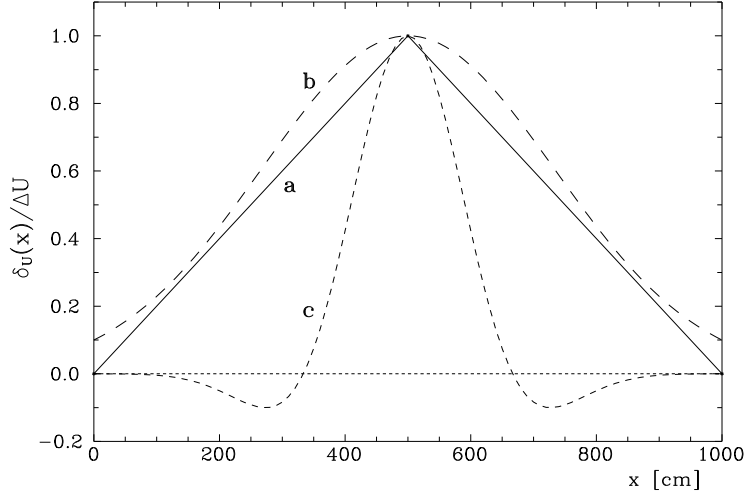}}
\caption{Shapes of disturbing potentials used in this work.
 (a) Triangular shape, (b) Gaussian one with $\varepsilon=0.1$, (c)
Alternating Quenching Gaussian with $\varepsilon=0.1$.}
\label{f:deltapot}
\end{figure}

The shape of the disturbing potential, $\delta_U(x)$, is not
known. Then it is necessary to use some reasonable estimate. We
tried three different shapes (see Fig.(\ref{f:deltapot})).
\begin{description}
\item{\bf Triangular shape}: the potential linearly decreases from
the value in the tritium injection point (in the middle of WGTS),
$\delta_U(L/2) = \Delta U$, to zero at both WGTS ends
\beq
    \delta_U(x) = \left\{
              \begin{array}{ll}
              2\Delta U/L\times x, & \hspace{2em} 0\leq x\leq L/2\\
              2\Delta U(1-x/L), & \hspace{2em} L/2 < x.
              \end{array}
                  \right.
\eeq \vspace{0.3ex}
\item{\bf Gaussian shape}: again, the potential decreases from
$\Delta U$ but to some small value of $\delta_U(0) = \delta_U(L) =
\varepsilon\times\Delta U$ at both ends and the shape is Gaussian
instead of linear one
\beq
     \delta_U(x) = \Delta U\times \exp{\{-\alpha(x-L/2)^2\}},
\eeq
where $\alpha = -4\ln{\varepsilon}/L^2$ and its FWHM is
$2\sqrt{\ln{2}/\alpha}$.  \vspace{1ex}
\item{\bf Alternating Quenching Gaussian}: this is Gaussian
modulated by a periodical function. It gives regions of small
potential with opposite sign near the both WGTS ends
\beq
    \delta_U(x) = \Delta U\times
    \exp{\{-\alpha(x-L/2)^2\}}\cos{\beta(x-L/2)}.
\eeq
We request the minimum being (in absolute value) a small part of
the maximum value, $\delta_U(x_{min}) = -\varepsilon\times\Delta
U$ and the potential to diminish at both ends. These two
conditions lead to a system of transcendental equations for \alf,
\bet, and $x_{min}$. Fortunately, $\beta = 3\pi/L$ may be
extracted and \alf\ and $x_{min}$ may be separated. We get the
equation $2\ln{\frac{\cos{\beta x_{min}}}{\varepsilon}} + \beta
x_{min}\times \tan{\beta x_{min}} = 0$ which can be solved
numerically and $\alpha = -\beta/(2x_{min})\times\tan{\beta
x_{min}}$.

\end{description}
\section{Numerics}
The simulated spectra of the \ikry\ L\dn{3} conversion line were
calculated in the energy interval from 30471~eV to 30475.4~eV with
the step of 0.1~eV, i.e. in 45 points. The quantities
characteristic for the experimental setup were taken those
expected \cite{DesRep} to be used for KATRIN. In particular
$B_{max}$=6~T, $B_A$=3~G, $B_s$=3.6~T, and $L$=10~m. For the
Gaussian and Alternating Quenching Gaussian shapes,
$\varepsilon$=10\up{-1} was chosen in both cases.

The tritium density, $\rho(x)$ is assumed to copy the tritium
pressure dependence which we took following to Sharipov
\cite{Shari}. The probability $p_0(x)$ was calculated by Ka\v spar
\cite{Kaspriv}.

To calculate the integrals in (\ref{e:diff}), we utilized the
Gauss-Legendre method as implemented in \cite{Ry84}. The
subsequent convolution with the instrumental function was done by
simple summation using a suitably small step. The relative
precision was checked to be better than 10\up{-3} which is quite
sufficient for our purpose. The calculated spectra were then
randomized in accordance with corresponding normal distribution.

The tests were performed as follows: For a particular set of
activity $A$, exposure time, and disturbing potential type eight
spectra were generated corresponding to different $\Delta U$'s.
Then they were compared, each to each, using both $\chi^2$ and
Student tests in both ratio and difference methods \cite{Dra97}.
(Since all four tests gave comparable results we shall present
only those of the $\chi^2$ test of the ratio method.) The
procedure was repeated ten times thus supplying a set of the test
values -- $\chi^2$'s per degree of freedom, $\chi^2_\nu$, in this
case. As the resulting value of ``sensitivity'' we adopt such
difference $\Delta U_i - \Delta U_j$ for which most of the
$\chi^2_\nu$ values are greater than 1.5 and no one less than 1.3.
\section{Results}
We have considered all three types of the disturbing potentials
described. In every case, the exposure time of 1 min in point,
i.e. 45 min per spectrum, was assumed. The calculations were done
for the \ikry\ activities of 10, 20, 50, 100, and 200~kBq,
respectively, and counts registered by the whole detector were
included.
\begin{table}[h]
 \caption{Intervals of the $\chi^2_\nu$ values
obtained during simulations. Activity od 20~kBq in WGTS, exposure
time of 1~min in point, and detection by the whole detector are
assumed.} \vspace{2mm} \label{t:region}
\begin{tabular}{|c|ccc|}
\hline
 simulation & \multicolumn{3}{c|}{$\Delta U_i - \Delta U_j \left[V\right]$} \\
\cline{2-4} 
number & 0.02 & 0.03 & 0.04 \\
\hline
1  & 1.1 -- 1.4 & 1.1 -- 1.9 & 1.6 -- 2.1\\
2  & 1.2 -- 1.5 & 1.3 -- 1.9 & 2.0 -- 2.2\\
3  & 0.8 -- 1.7 & 1.7 -- 1.9 & 1.3 -- 2.2\\
4  & 0.8 -- 1.9 & 1.1 -- 2.1 & 1.4 -- 2.1\\
5  & 0.8 -- 1.7 & 1.0 -- 2.0 & 1.7 -- 2.7\\
6  & 0.9 -- 1.6 & 1.3 -- 2.0 & 1.8 -- 2.3\\
7  & 0.9 -- 1.6 & 1.3 -- 2.1 & 1.9 -- 2.7\\
8  & 1.0 -- 1.3 & 1.1 -- 1.9 & 1.5 -- 2.2\\
9  & 1.2 -- 1.7 & 1.4 -- 2.1 & 2.0 -- 2.6\\
10 & 0.9 -- 1.3 & 1.0 -- 1.8 & 1.7 -- 2.0\\
 \hline
\end{tabular}
\end{table}

\begin{figure}[h]
  \centerline{\includegraphics[clip=on,angle=0,width=10cm]{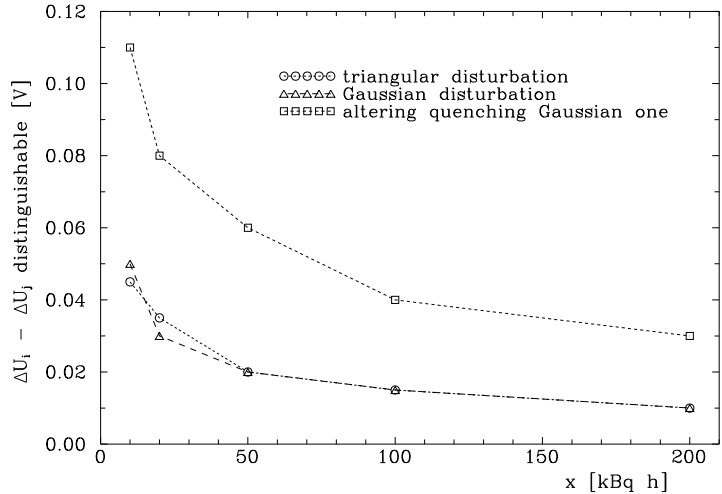}}
\caption{Sensitivities found for the three types of disturbing
potential studied in this work. For the meaning of the variable
$x$ see text.} \label{f:sens}
\end{figure}

As mentioned above the spectra for every particular setup were
simulated ten times, each comparison giving an
interval\footnote{E.g. for $\Delta U_i - \Delta U_j = 0.02$, we
compare pairs of spectra for particular $\Delta U$'s (0,0.02),
(0.01,0.03), (0.02,0.04) etc. We then present only the minimum and
maximum of $\chi^2_\nu$.} of the $\chi^2_\nu$ values. For an
illustration, one such example is shown in Table \ref{t:region}.
In that case it is clear that the sensitivity of 40~mV is assured
but, moreover, that of 35~mV can be expected safely enough.
The sensitivities deduced in this way are presented in Fig.
\ref{f:sens}. Since we do not know exactly the shape of the
disturbing potential it would be the most reasonable to use the
worst (highest) values as estimates.

It is necessary to remind here that the results are determined by
the {\em total number of counts registered}. This means that the
result for e.g. the activity $A$ and exposure $T$ are the same as
those for the activity of $A/2$ and exposure of $2T$. Similar
argument holds when we compare the spectra measured by various
parts of the detector. This is expressed in Fig. \ref{f:sens} by
using at the horizontal axis the quantity, $x$, which is defined
as
\beq
   x = \frac{n}{n_{tot}}\times A\times\frac{4}{3}T,
\eeq
where $n_{tot}$ is the total number of pixels in the detector, $n$
is the number of active (registering) pixels, $A$ is the \ikry\
activity within WGTS in kBq, and $T$ is measurement time (of the
whole spectrum) in hours.\\
%

\section*{Acknoledgement}
This work was partly supported by the Grant Agency of the Czech
   Republic under contract No. 202/02/0889 and by the ASCR project AV0Z10480505
%

%
%
%
\end{document}